\documentclass[
    ,final            
  ]
  {aipproc}

\layoutstyle{8x11single}

\begin{document}

\title{The helium star donor channel for the progenitors of type Ia supernovae and their surviving companion stars}


\author{Bo Wang}
{
  address={National Astronomical Observatories/Yunnan Observatory, CAS, Kunming, 650011, China; {\it wangbo@ynao.ac.cn, zhanwenhan@ynao.ac.cn}}
  ,altaddress={Key Laboratory for the Structure and Evolution of Celestial Objects, CAS, Kunming 650011, China} 
}

\author{Zhanwen Han}
{
  address={National Astronomical Observatories/Yunnan Observatory, CAS, Kunming, 650011, China; {\it wangbo@ynao.ac.cn, zhanwenhan@ynao.ac.cn}}
  ,altaddress={Key Laboratory for the Structure and Evolution of Celestial Objects, CAS, Kunming 650011, China} 
}


\begin{abstract}
The nature of type Ia supernovae (SNe Ia) is still unclear.
Employing Eggleton's stellar evolution code with the optically thick
wind assumption, we systematically studied the He star donor channel
of SNe Ia, in which a carbon-oxygen white dwarf accretes material
from a He main-sequence star or a He subgiant to increase its mass
to the Chandrasekhar mass. We mapped out the initial parameters for
producing SNe Ia in the orbital period--secondary mass plane for
various WD masses from this channel. According to a detailed binary
population synthesis approach, we find that this channel can produce
SNe Ia with short delay times ($\sim$100\,Myr) implied by recent
observations. We obtained many properties of the surviving
companions of this channel after SN explosion, which can be verified
by future observations. We also find that the surviving companions
from the SN explosion scenario have a high spatial velocity
($>$400\,km/s), which could be an alternative origin for
hypervelocity stars (HVSs), especially for HVSs such as US 708.
\end{abstract}

\maketitle


\section{1. Introduction}
\label{sect:intro}

Type Ia supernovae (SNe Ia) play an important role in astrophysics,
especially in the study of cosmic evolution. They have been applied
successfully in determining cosmological parameters (e.g.,
\textbf{$\Omega_{M}$} and \textbf{$\Omega_{\Lambda}$}; \cite{rie98,
per99}). They are also a key part of our understanding of galactic
chemical evolution owing to the main contribution of iron to their
host galaxies. It is generally believed that SNe Ia are
thermonuclear explosions of carbon-oxygen white dwarfs (CO WDs) in
binaries. However, there is still no agreement on the nature of
their progenitors [3$-$6], and no SN Ia progenitor system before the
explosion has been conclusively identified [7$-$9].

Over the past few decades, two families of SN Ia progenitor models
have been proposed, i.e., the double-degenerate (DD) and
single-degenerate (SD) models. Of these two models, the SD model is
widely accepted at present. It is suggested that the DD model, which
involves the merger of two CO WDs [10$-$12], likely leads to an
accretion-induced collapse rather than to an SN Ia \cite{nom85}. For
the SD model, the companion could be a main-sequence (MS) star or a
slightly evolved subgiant star (WD + MS channel), or a red-giant
star (WD + RG channel) [14$-$26]. Observationally, there is
increasing evidence indicating that at least some SNe Ia may come
from the SD model [27$-$30].

Meanwhile, a CO WD may also accrete material from a He star to
increase its mass to the Chandrasekhar (Ch) mass, which is known as
the He star donor channel. Yoon and Langer \cite{yl03} followed the
evolution of a CO WD + He star system, in which the WD can increase
its mass to the Ch mass by accreting material from the He star. It
is believed that WD + He star systems are generally originated from
intermediate mass binaries, which may explain SNe Ia with short
delay times implied by recent observations [32$-$35].

In this contribution, we aim to investigate SN Ia birthrates and
delay times of the He star donor channel and to explore the
properties of the surviving companions after SN explosion. In Sect.
2, we describe the numerical code for the binary evolution
calculations and the binary evolutionary results. We describe the
binary population synthesis (BPS) method and results in Sect. 3.
Finally, a discussion is given in Sect. 4.

\section{2. Binary evolution calculations}

\begin{figure}[tb]
\includegraphics[width=6.5cm,angle=270]{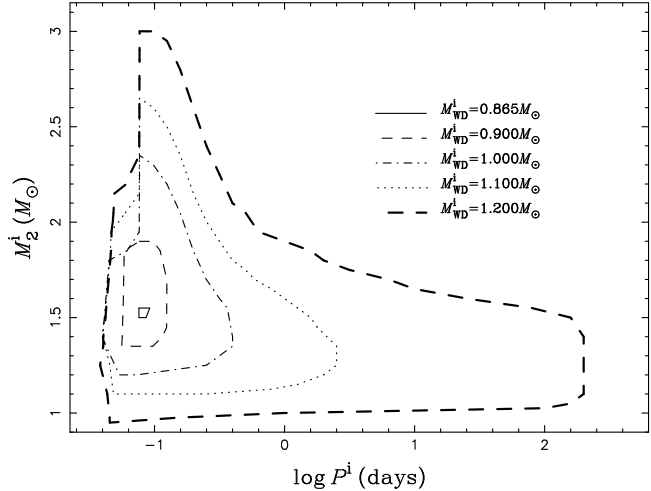}
 \caption{Regions in the orbital
period--secondary mass plane for WD binaries (at the onset of the
RLOF stage) that produce SNe Ia for initial WD masses of $0.865,
0.9, 1.0, 1.1$ and $1.2$\,$M_{\odot}$. (From Wang et al.
\cite{wan09a})}
\end{figure}

We use Eggleton's stellar evolution code \cite{egg71, egg72, egg73}
to calculate the binary evolutions of WD + He star systems. The code
has been updated with the latest input physics over the past three
decades \cite{han94, pol95, pol98}. Roche lobe overflow (RLOF) is
treated within the code described by Han et al. \cite{han00}. We set
the ratio of mixing length to local pressure scale height,
$\alpha=l/H_{\rm p}$, to be 2.0. The opacity tables are compiled by
Chen and Tout \cite{ct07}. In our calculations, He star models are
composed by He abundance $Y=0.98$ and metallicity $Z=0.02$.

Instead of solving stellar structure equations of a WD, we use the
optically thick wind model \cite{hac96} and adopt the prescription
of Kato \& Hachisu \cite{kh04} for the mass accumulation efficiency
of He-shell flashes onto the WD. We have calculated about 2600 WD +
He star systems, and obtain a large, dense model grid (for details
see \cite{wan09a}). In Fig. 1, we show the contours at the onset of
RLOF for producing SNe Ia in the $\log P^{\rm i}-M^{\rm i}_2$ plane
for various WD masses (i.e., $M_{\rm WD}^{\rm i}$ = 0.865, 0.90,
1.0, 1.1 and $1.2\,M_{\odot}$), where $P^{\rm i}$ and $M^{\rm i}_2$
are the orbital period and the mass of the He companion star at the
onset of RLOF, respectively. The region almost vanishes for $M_{\rm
WD}^{\rm i}$=0.865\,$M_{\odot}$, which is then assumed to be the
minimum WD mass for producing SNe Ia from this channel. If the
parameters of a WD + He star system at the onset of the RLOF are
located in the contours, an SN Ia is then assumed to be produced.
Thus, these contours can be expediently used in BPS studies to
investigate the He star donor channel of SNe Ia. The data points and
the interpolation FORTRAN code for these contours can be supplied on
request by contacting BW.

\section{3. Binary population synthesis}
\begin{figure}
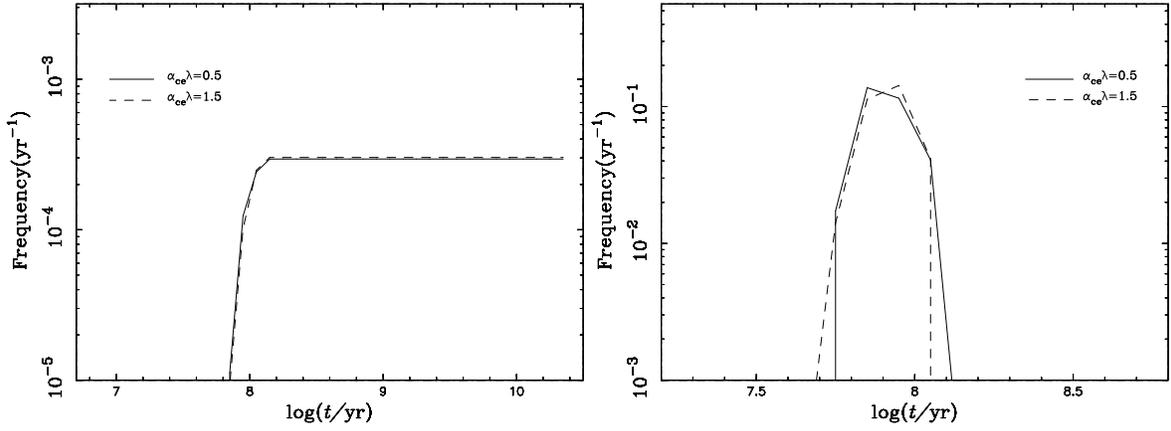

\includegraphics[width=5.6cm,angle=270]{f2a.ps}
\includegraphics[width=5.6cm,angle=270]{f2b.ps}
 \caption{The evolution of SN Ia birthrates with time for the He star donor channel.
\textit{Left panel}: the evolution of Galactic SN Ia birthrates for
a constant Pop I SFR ($5\,M_{\odot}$\,yr$^{-1}$), where the solid
and dashed curves show the results of different CE ejection
parameters with $\alpha_{\rm ce}\lambda=0.5$ (solid) and
$\alpha_{\rm ce}\lambda=1.5$ (dashed), respectively. \textit{Right
panel}: similar to the left panel, but for a single starburst with a
total mass of $10^{\rm 11}\,M_{\odot}$. (From Wang et al.
\cite{wan09b})}
\end{figure}

To obtain SN Ia birthrates and delay times of the He star donor
channel, we performed a series of Monte Carlo simulations in the BPS
study. In the simulation, by using the Hurley's rapid binary
evolution code \cite{hur00, hur02}, we followed the evolution of
$4\times10^{\rm 7}$ sample binaries from the star formation to the
formation of the WD + He star systems according to the SN Ia
production regions (Fig. 1) and three evolutionary channels (i.e.,
the \textit{He star channel}, the \textit{EAGB channel}, and the
\textit{TPAGB channel}; for details see \cite{wan09b}). Here, we
adopt the standard energy equations to calculate the output of the
common envelop (CE) phase (see \cite{wan09b}).

In the BPS study, the primordial binary samples are generated in the
Monte Carlo way and a circular orbit is assumed for all binaries. We
adopt the following input for the simulation (for details see
\cite{wan09b}). (1) The initial mass function of Miller and Scalo
\cite{ms79} is adopted. (2) The mass-ratio distribution is taken to
be constant. (3) The distribution of separations is taken to be
constant in $\log a$ for wide binaries, where $a$ is the orbital
separation. (4) We simply assume a constant star formation rate
(SFR) over the past 15\,Gyr or, alternatively, as a delta function,
i.e., a single starburst.

In Fig. 2, we show the evolution of SN Ia birthrates with time for
the He star donor channel. The left panel represents Galactic
birthrates of SNe Ia by adopting $Z=0.02$ and ${\rm SFR}=5\,M_{\rm
\odot}{\rm yr}^{-1}$. The simulations give Galactic SN Ia birthrates
of $\sim$$0.3\times 10^{-3}\ {\rm yr}^{-1}$. The right panel
displays the evolution of SN Ia birthrates for a single starburst
with a total mass of $10^{11}\,M_{\odot}$. In this panel, we see
that SN Ia explosions occur between $\sim$45\,Myr and $\sim$
140\,Myr after the starburst, which could explain SNe Ia with short
delay times implied by recent observations [32$-$35].

\begin{figure}
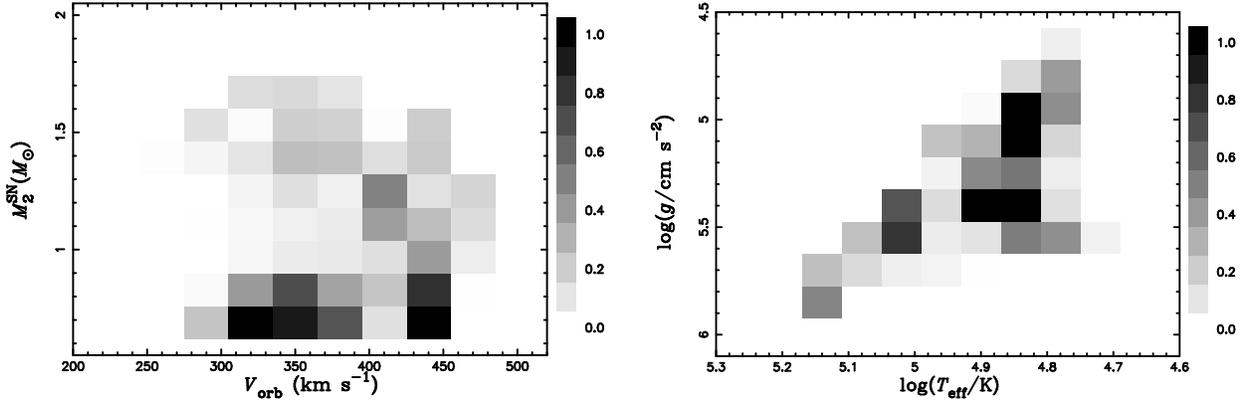

\includegraphics[width=5.3cm,angle=270]{f3a.ps}
\includegraphics[width=5.3cm,angle=270]{f3b.ps}
 \caption{\textit{Left panel}:
the distribution of properties of companions in the plane of
($V_{\rm orb}$, $M_2^{\rm SN}$) at the current epoch, where $V_{\rm
orb}$ is the orbital velocity  and $M_2^{\rm SN}$ the mass at the
moment of SN explosion. \textit{Right panel}: similar to the left
panel, but in the plane of ($\log T_{\rm eff}$, $\log g$), where
$T_{\rm eff}$ is the effective temperature of companions and $\log
g$ the surface gravity. (From Wang \& Han \cite{wh09a})}
\end{figure}

\begin{figure}
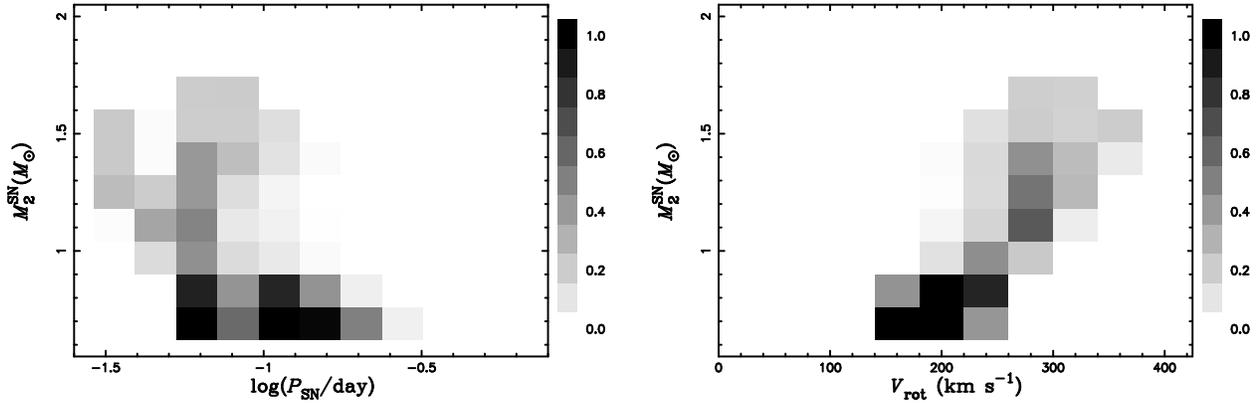

\includegraphics[width=5.3cm,angle=270]{f4a.ps}
\includegraphics[width=5.3cm,angle=270]{f4b.ps}
 \caption{\textit{Left panel}:
similar to Fig. 3, but in the plane of ($\log P^{\rm SN}$, $M_2^{\rm
SN}$), where $P^{\rm SN}$is the orbital period. \textit{Right
panel}: similar to Fig. 3, but in the plane of ($V_{\rm rot}$,
$M_2^{\rm SN}$), where $V_{\rm rot}$ is the equatorial rotational
velocity of companions. (From Wang \& Han \cite{wh09a})}
\end{figure}

The companion in the SD model would survive in the SN explosion and
potentially be identifiable \cite{wh10c, pan10}. We obtained the
distributions of many properties of the surviving companions of this
channel at the moment of SN explosion. Figs 3 and 4 shows the
distributions of the masses, the orbital velocities, the effective
temperatures, the surface gravities, the orbital period and the
equatorial rotational velocity of companions at the moment of SN
explosion (for the detailed discussion see \cite{wh09a}). These
distributions may be helpful for identifying the surviving
companions of SNe Ia. The simulation also shows the initial
parameters of the primordial binaries and the WD + He star systems
that lead to SNe Ia, which may help to search for potential SN Ia
progenitors.

\begin{figure}[tb]
\includegraphics[width=8.5cm,angle=0]{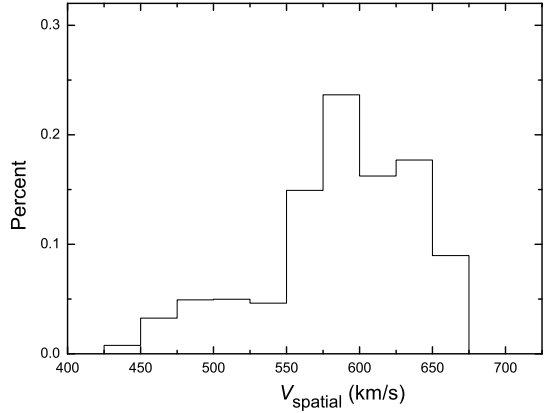}
 \caption{The distribution of the spatial velocity with
 $\alpha_{\rm ce}\lambda=0.5$. (From Wang \& Han \cite{wh09a})}
\end{figure}

In Fig. 5, we show the current epoch distribution of the spatial
velocity for the surviving companions from this channel (for details
see \cite{wh09a}). We see that the surviving companions have high
spatial velocities ($>$400\,km/s), which almost exceed the
gravitational pull of the Galaxy nearby the Sun. Thus, the surviving
companions from the SN explosion scenario could be an alternative
origin for hypervelocity stars (HVSs), which are stars with a
velocity so great that they are able to escape the gravitational
pull of the Galaxy.

\section{4. Discussion}
The simulations give Galactic SN Ia birthrate of $\sim$$0.3\times
10^{-3}\ {\rm yr}^{-1}$ for the He star donor channel, which is
lower than that inferred observationally (i.e., $3 - 4\times
10^{-3}\ {\rm yr}^{-1}$; Cappellaro and Turatto \cite{ct97}). This
implies that the He star donor channel is only a subclass of SN Ia
production, and there may be some other channels or mechanisms also
contributing to SNe Ia, e.g., WD + MS channel, WD + RG channel or
double-degenerate channel (see \cite{wan10}).

Hachisu et al. \cite{hac08} recently studied the mass-stripping
effect on a MS or slightly evolved companion star by winds from a
mass-accreting WD. The model can also provide a possible ways of
producing young SNe Ia, but the model depends on the efficiency of
the mass-stripping effect. We also find that the model produces very
few young SNe Ia according to a detailed BPS approach. Thus, we
consider the He star donor channel as a main contribution to the
formation of young SNe Ia.

Based on the optically thick wind assumption, Wang \& Han
\cite{wh10d} recently calculated about 10 000 WD + He star systems
and obtained SN Ia production regions of the He star donor channel
with different metallicities. For a constant star-formation galaxy,
they found that SN Ia birthrates increase with metallicity. If a
single starburst is assumed, SNe Ia occur systemically earlier and
the peak value of the birthrate is larger for a high $Z$. We also
note that Liu et al. \cite{liu10} recently investigated the effects
of rapid differential rotation on the accreting WD, and found that
the highest mass of the accreting WD at the moment of SN Ia
explosion is $1.81\,M_{\odot}$ for the He star donor channel, which
may provide a way for the formation of super-Ch mass SNe Ia in the
observations.

\begin{figure}
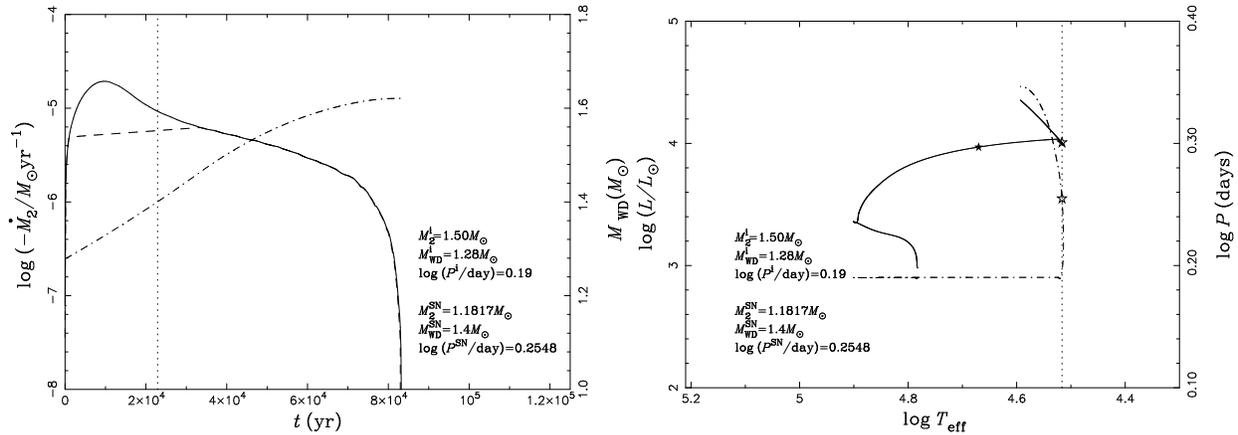

\includegraphics[width=5.7cm,angle=270]{f6a.ps}
\includegraphics[width=5.7cm,angle=270]{f6b.ps}
 \caption{Results of binary evolution calculations with initial masses of two components and
the orbital period as well similar to the binary system HD49798/RX
J0648.9-4418. In the left panel, the solid, dashed and dash-dotted
curves show the mass-transfer rate, the mass-growth rate of the WD
and the mass of the WD varying with time after HD 49798 fills its
Roche lobe, respectively. In the right panel, the evolutionary track
of the He donor star is shown as a solid curve, and the evolution of
the orbital period is shown as a dash-dotted curve. The solid star
represents the current position of HD 49798. Dotted vertical lines
in these two panels and open stars in the right panel indicate the
position where the WD is expected to explode as an SN Ia. (From Wang
\& Han \cite{wh10e})}
\end{figure}

Massive WD + He star systems are candidates of SN Ia progenitors. HD
49798/RX J0648.0-4418 is an evidence of the existence of massive WD
+ He star systems. Based on the data from the XMM-Newton satellite,
Mereghetti et al. \cite{mer09} recently derived the masses of the
two components. The corresponding masses are
1.50$\pm$0.05$\,M_{\odot}$ for HD 49798 and
1.28$\pm$0.05$\,M_{\odot}$ for the WD. According to a detailed
binary evolution model, Wang \& Han \cite{wh10e} found that the
massive WD can increase its mass to the Ch mass in future evolution
(see Fig. 6). Thus, HD 49798 with its WD companion is a likely
candidate of SN Ia progenitors. V445 Pup is another candidate of
massive WD + He star system, in which the mass of the WD is more
than $1.35\,M_{\odot}$, and the mass of the He star is more than
$0.85\,M_{\odot}$ \cite{kat08, wou09}. However, we still do not know
the orbital period of the binary system and the mass of the He donor
star so far. This needs further observations of V445 Pup after the
dense dust shell disappears.

US 708 is an extremely He-rich sdO star in the Galaxy halo, with a
heliocentric radial velocity of +$708\pm15$\,${\rm km/s}$ (Hirsch et
al. \cite{hir05}). We note that the local velocity relative to the
Galatic center may lead to a higher observation velocity for the
surviving companions, but this may also lead to a lower observation
velocity. Considering the local velocity nearby the Sun
($\sim$220\,km/s), we find that $\sim$30\% of the surviving
companions may be observed to have velocity $V>700\,{\rm km/s}$ for
a given SN ejecta velocity 13500\,km/s. In addition, the asymmetric
explosion of SNe Ia may also enhance the velocity of the surviving
companions. Thus, a surviving companion in the He star donor channel
may have a high velocity like US 708. In future investigations, we
will employ the Large sky Area Multi-Object fiber Spectral Telescope
(LAMOST) to search the HVSs originating from the surviving
companions of SNe Ia.


\begin{theacknowledgments}
This work is supported by the National Natural Science Foundation of
China (Grant Nos. 10821061 and  11033008), the National Basic
Research Program of China (Grant No. 2007CB815406) and the Chinese
Academy of Sciences (Grant No. KJCX2-YW-T24).
\end{theacknowledgments}

\bibliographystyle{aipprocl} 

\end{document}